%% file: main.tex
\documentclass{article}

\usepackage[latin1]{inputenc}
\usepackage{graphicx}

\usepackage{amsmath}

\usepackage{algorithmic}
\usepackage{algorithm}

\usepackage{amsfonts}
\usepackage{color}

\providecommand{\norm}[1]{\lVert#1\rVert}
\providecommand{\abs}[1]{\lvert#1\rvert}
\providecommand{\ind}[1]{\mathbf{1}_{#1}}

\newcommand{\Prob}{\mathbb{P}}
\newcommand{\R}{\mathbb{R}}
\newcommand{\E}{\mathbb{E}}
\newcommand{\gk}{\Gamma_{\!k}}
\newcommand{\gkk}{\Gamma_{\!k+\!1}}
\newcommand{\gkm}{\Gamma_{\!k-\!1}}
\newcommand{\vqXk}{\widehat X_k^{\gk}}
\newcommand{\vqXkk}{\widehat X_{k+1}^{\gkk}}
\newcommand{\vqXkm}{\widehat X_{k-1}^{\gkm}}
\newcommand{\vqXn}{\widehat X_n^{\Gamma_{\!n}}}
\newcommand{\qXk}{\hat X_k}
\newcommand{\qXkk}{\hat X_{k+1}}

\newcommand{\tp}{\pi^k_{ij}}
\newcommand{\Qn}{Q_{\text{min}}}
\newcommand{\Qx}{Q_{\text{max}}}

\newcommand{\cuda}{{\tt CUDA}}
\newcommand{\nvidia}{NVIDIA}
\newcommand{\gpgpu}{{\tt GPGPU}}
\newcommand{\CCone}{\cuda\ Compute Capability 1.x}
\newcommand{\CCtwo}{\cuda\ Compute Capability 2.x}
\newcommand{\FermiD}{{\tt NVIDIA GTX 480}}
\newcommand{\oldD}{{\tt NVIDIA GTX 295}}

\DeclareMathOperator{\esup}{esssup}

\newtheorem{prop}{Proposition}

\begin{document}


\input{title}
\maketitle

\begin{abstract}
The pricing of American style and multiple exercise options is a very challenging problem in mathematical finance. 
One usually employs a Least-Square Monte Carlo approach (Longstaff-Schwartz method) for the evaluation of conditional expectations which arise in the Backward Dynamic Programming principle for such optimal stopping or stochastic control problems in a Markovian framework.
Unfortunately, these Least-Square Monte Carlo approaches are rather slow and allow, due to the dependency structure in the Backward Dynamic Programming principle, no parallel implementation; whether on the Monte Carlo level nor on the time layer level of this problem.

We therefore present in this paper a quantization method for the computation of the conditional expectations, that allows a straightforward parallelization on the Monte Carlo level.
Moreover, we are able to develop for AR(1)-processes a further parallelization in the time domain, which makes use of faster memory structures and therefore maximizes parallel execution.

Finally, we present numerical results for a CUDA implementation of this methods. It will turn out 
that such an implementation leads to an impressive speed-up compared to a serial CPU implementation.

\end{abstract}


\input{paper}

\bibliographystyle{plain}

\bibliography{myliterature}

\end{document}

%% file: title.tex
\title{GPGPUs in computational finance: Massive parallel computing for American style options}

\author{Gilles Pagès\thanks{Laboratoire de Probabilit\'es et Mod\`eles al\'eatoires, UMR~7599, Universit\'e Paris 6, case 188, 4, pl. Jussieu, F-75252 Paris Cedex 05. E-mail: {\tt  gilles.pages@upmc.fr}}
 \and Benedikt Wilbertz\thanks{Laboratoire de Probabilit\'es et Mod\`eles al\'eatoires, UMR~7599, Universit\'e Paris 6, case 188, 4, pl. Jussieu, F-75252 Paris Cedex 05. E-mail: {\tt  benedikt.wilbertz@upmc.fr}}
}


%% file: paper.tex
\section{Introduction}

The pricing of American style and multiple exercise options consists of solving the optimal stopping problem
\[
  V = \esup \Bigl\{ \E\bigl( \varphi_\tau(X_{\tau}) \bigl| \mathcal{F}_0 \bigr): \tau \text{ is a }
(\mathcal{F}_k)
\text{-stopping time}
\Bigr\} 
\]
for an adapted stochastic process $(X_k)_{0\leq k \leq n}$ on a filtered probability space  $(\Omega, (\mathcal{F}_k)_{0\leq k \leq n}, \Prob)$ and obstacle functionals $\varphi_k, 0\leq k \leq n$.

It is well known (see e.g. \cite{snell})
that $V$ is given by the solution $V_0$ to the  {\it Backward Dynamic Programming (BDP) Principle} 
\begin{equation}\label{eqBDP}
  \begin{split}
    V_n & = \varphi_{t_n}(X_n)\\
    V_k & = \max \Bigl( \varphi_{t_k}(X_k); \, \E\bigl( V_{k+1} \bigl| \mathcal{F}_k\bigr) \Bigr), \,\, 0 \leq k \leq n-1.
  \end{split}
\end{equation}

We focus here on the case of an adapted Markov chain $(X_k)$, so that it holds $\E( V_{k+1} | \mathcal{F}_k) = \E( V_{k+1} | X_k) $.
Then the main difficulty of solving (\ref{eqBDP}) by means of Monte Carlo methods lies in the approximation of the conditional expectations $\E( V_{k+1} | X_k)$.
This is usually accomplished by a Least Squares regression as proposed by the Longstaff-Schwartz method.
Following \cite{LScarriere,LS} and \cite{vanRoy} the main steps of this procedure consists of
\begin{itemize}
\item Simulating $M$ paths of $(X_k)$ (forward step)
\item Starting at $k = n-1$, approximate $f_k(x) = \E( V_{k+1} | X_k = x)$ by a Least Squares regression and proceed backwards to 0. (backward step)
\end{itemize}

From a practical point of view, the most expensive tasks are clearly the repeated Least Square regressions on the huge number of Monte Carlo paths.
Due to the sequential dependency structure of the Backward Dynamic Programming formula, the collection of the Least Squares problems as a whole cannot be solved in parallel, but has to be processed in strict sequence.
Moreover, it is not an easy task to solve the single Least Square problems efficiently in parallel.

We therefore present in this paper a Quantization Tree algorithm, which handles the most part of the work in a forward step which can be easily parallelized on the Monte Carlo level (pathwise) as well as on the time layer level. 
Therefore, this approach is well suited for the use of massive parallel computing devices like GPGPUs.
Using this approach, the subsequent backward processing of the BDP principle becomes straightforward and negligible in terms of computational costs when compared to the Least Squares backward step.

\section{The Quantization Tree Algorithm}
The Quantization Tree algorithm is an efficient tool to establish a pathwise discretization of a discrete-time Markov chain (see e.g. \cite{ballyPages, BPP, BBP1} or \cite{BPW}).
Such a discretization can be used to solve optimal stopping or control problems, as they occur in the evaluation of financial derivatives with non-vanilla exercise rights.
In this paper, we focus on a fast computation of the transition probabilities in a Quantization Tree by means of \gpgpu-devices, which make this approach suitable for time-critical online computations.

Therefore, let $(X_k)_{0\leq k\leq n}$ be a discrete-time $L^2$-Markov chain on a filtered probability space $(\Omega, (\mathcal{F}_k)_{0\leq k \leq n}, \Prob)$ with values in the vector space $(\R^d, \mathcal{B}^d)$.
This vector space shall be endowed with an appropriated norm (often Euclidean norm).
For each time-step $k$ we furthermore assume to a have a quantization grid
\[
\gk = (x_1^k, \ldots, x_{N_k}^k)
\]
of size $N_k$.

This means that $\gk$ provides a discretization of the state space of the r.v. $X_k$, which is supposed to minimize the quadratic quantization error
\begin{equation}
  \E \min_{1\leq i\leq N_k} \norm{X_k - x_i^k}^2
\label{eq:Qerr}
\end{equation}
over all possible grids $\gk\subset \R^d$ with size $\abs{\gk}\leq N_k$. (See \cite{Foundations} for a comprehensive introduction to quantization of probability distributions.)

For a grid $\gk$, let $\bigl(C_i(\gk) \bigr)_{1\leq i \leq N_k}$ be a {\it Voronoi Partition} of $\R^d$ induced by the points in $\gk$, i.e.
\[
C_i(\gk) \subset \bigl\{y \in \R^d: \norm{y-x_i^k} \leq \min_{1\leq j \leq N_k} \norm{y-x_j^k}  \bigr\}.
\]

We then call the mapping
\[
 z \mapsto \sum_{i=1}^{N_k} x_i \ind{C_i(\gk)}(z)
\]
the {\it Nearest Neighbor projection} of $z$ onto $\gk$.

This Nearest Neighbor projection defines in a natural way the {\it Voronoi Quantization}
\[
\vqXk = \sum_{i=1}^{N_k} x_i \ind{C_i(\gk)}(X_k),
\]
which obviously provides a discrete r.v. with not more than $N_k$ states and
\[
\E\norm{X_k - \vqXk}^2 = \E \min_{1\leq i \leq N_k} \norm{X_k - x_i^k}^2.
\]

\begin{figure}[!t]
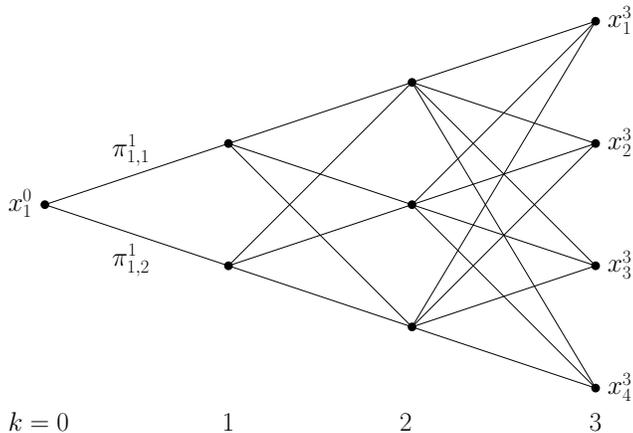

\centering
\resizebox{8cm}{!}{\input tree_t}
\caption{A Quantization Tree $\Gamma$}
\label{fig_tree}
\end{figure}

Defining the cartesian product quantizer
\[
\Gamma = \prod_{k=0}^n \gk
\]
we arrive at a path discretization of the Markov chain $(X_k)$ with $\abs{\Gamma} \leq \prod_{k=0}^n N_k$ paths, which we will call
%
the {\it Quantization Tree} (see Figure \ref{fig_tree}).

To equip $\Gamma$ with a probability distribution, we introduce the transition probabilities
\begin{equation}
  \begin{split}
 \tp & = \Prob\bigl(\vqXk = x_j^{k}\, |\, \vqXkm = x_i^{k-1} \bigr)\\
   & = \Prob\bigl( X_{k} \in C_j(\gk)\, |\, X_{k-1} \in C_i(\gkm)  \bigr).
\end{split}\label{eq:defPi}
\end{equation}

If the marginal distributions of $(X_k)$ are Gaussian and the norm is the canonical Euclidean norm, grids which minimize (\ref{eq:Qerr}) are precomputed and available at \cite{Website}.
Otherwise, some sub-optimal grids, matching the first two moments of $X_k$, can be employed at the price of not achieving the full optimal convergence rate.

Nevertheless, the true difficulties of this approach actually consist in the computations of the transition probabilities $\tp$.
These probabilities are usually so strongly connected to the individual choice of the Markov chain $(X_k)$ that they cannot be precomputed like the above quantization grids or approximated by simple means.

We therefore have to perform a Monte-Carlo (MC) simulation of the Markov chain $(X_k)$ in order to estimate the transition probabilities $\tp$.
%
%
%
Since these MC simulations can be quite time consuming, we will take advantage of the massive parallel computing capabilities of nowadays \gpgpu-devices 
and reduce the computational time for the estimation of the transition probabilities to a level that actually is acceptable for time-critical applications in financial practice.

As the Quantization Tree $\Gamma$ exhibits a pathwise approximation of the Markov chain $(X_k)$,
we may numerically solve on $\Gamma$ stochastic control or optimal stopping problems like they occur e.g. in the valuation of options with non-vanilla right exercises.

In \cite{BPP}, the optimal stopping problem

\begin{equation}
  \label{eq:optStopp}
  V = \esup \Bigl\{ \E\bigl( \varphi_\tau(X_{\tau}) \bigl| \mathcal{F}_0 \bigr): \tau \text{ a }
(\mathcal{F}_k)
\text{-stopping time}
\Bigr\} 
\end{equation}
with a payoff function $\varphi_t(x) = \bigl( s_0 \exp\bigl( (r-\sigma^2/2)t + \sigma x  \bigr) - K \bigr)^+$ and $(X_k)$ a $d$-dimensional time-discretized Brownian motion is solved to approximate American option prices.

In \cite{BBP1}, the authors employ the Quantization Tree to solve the stochastic control problem
\begin{equation}
  \label{eq:stochCtrl}
  \begin{split}
    P(Q)=\esup\biggl\{ \E \biggl( \sum_{k=0}^{n-1} q_k v_k(X_k) \Bigl|
    \mathcal{F}_0 \biggr)\!: \forall k = 0, \ldots, n\!-\!1:\\ 
    q_k\!: (\Omega, \mathcal{F}_k) \rightarrow
    [0,1], \,
 \sum_{k=0}^{n-1} q_k \in [\Qn, \Qx] \biggr\},
  \end{split}
\end{equation}
where $v_k$ can be interpreted as a payoff function and the couple $Q = (\Qn, \Qx)$ provides some global constraints on the cumulated consumption $\sum_{k=0}^{n-1} q_k$, so that (\ref{eq:stochCtrl}) yields the fair value of a swing option, which is an important derivative in energy trading.

Concerning the Quantization Tree algorithm, note that $\Gamma$ contains such a huge number of paths (e.g. at least $100^{365}$ in the example below) that it is impossible to process above problems in a path-wise manner.

Therefore, one usually resorts on the {\it Backward Dynamic Programming (BDP) Principle}, which allows a time-layer wise proceeding. This approach yields a complexity of  $C \sum_{k=1}^n N_{k-1} N_k$, i.e. increases only linearly in $n$.

In case of the optimal stopping problem (\ref{eq:optStopp}), the true BDP-principle 
can be approximated by setting
\begin{equation*}
  \begin{split}
    \widehat V_n & = \varphi_{t_n}(\vqXn)\\
    \widehat V_k & = \max \Bigl( \varphi_{t_k}(\vqXk); \, \E\bigl(\widehat V_{k+1} \bigl| \vqXk\bigr) \Bigr), \,\, 0 \leq k \leq n-1,
  \end{split}
\end{equation*}
so that the $\mathcal{F}_0$ measurable r.v. $\widehat V_0$ yields an approximation for $V$. 
Doing so we somehow ``force'' the Markov property of the Quantization sequence $(\vqXk)$.

In case of the stochastic control problem (\ref{eq:stochCtrl}), it was shown in \cite{BBP2} that there exists a bang-bang control for (\ref{eq:stochCtrl}), so that the BDP-principle leads to
\begin{equation*}
    \label{QTA}
    \begin{split}
      \widehat P_n &\equiv 0 \\
      \widehat P_k (Q^k) & = \max \Bigl\{ x v_k(\vqXk) \\ 
      & + \E\bigl(\widehat
        P_{k+1} (\chi^{n\!-\!k\!-\!1}(Q^k, x))\bigl|\vqXk\bigr), 
x \in \{0,1\} \cap
        I_{Q^k}^{n-k-1} \Bigr\},
    \end{split}
  \end{equation*}
where the set $I_{Q}^{k}$ and the function $\chi^{k}(Q, x)$ ensure to keep consumption within the global constraints $[\Qn, \Qx]$.

In both cases, we have to evaluate conditional expectations $\E\bigl(f(\qXkk)|\qXk\bigr)$, which reduce on $\Gamma$ to
\[
\E\bigl(f(\vqXkk)\bigl|\vqXk = x^k_i\bigl) = \sum_{j=1}^{N_{\!k\!+\!1}} f(x^{k\!+\!1}_j)\, \tp.
\]

Concerning the approximation error for this approach, assume that the $v_k$ are Lipschitz-continuous and that $(X_k)$ has Lipschitz-Feller transition kernels. We then get in case of a trivial $\sigma$-field $\mathcal{F}_0$ for a constant $C>0$ (see \cite{BBP2}, Thm 3)
\[
\abs{P(Q) - \widehat P_0(Q)} \leq C  \sum_{k=0}^{n-1} \Bigl( \E \norm{X_k - \vqXk}^2 \Bigr)^{1/2}.
\]

\section{Swing options in the Gaussian 2-factor model}
We will now focus on the implementation of the Quantization Tree algorithm for the valuation of Swing options in a Gaussian 2-factor model and present in detail the computation of the transition probabilities using \cuda\ on a \gpgpu-device.

In this model, the dynamics of the underlying are given as
\[
S_t = s_0 \exp\left(\sigma_1\!\! \int_0^{t} e^{-\alpha_1(t-s)} dW^1_s   + \sigma_2\!\! \int_0^{t} e^{-\alpha_2(t-s)} dW^2_s   -   \frac{1}{2} \mu_{t}\right)
\]
for Brownian Motions $W^1$ and $W^2$ with some correlation parameter $\rho$. 

Having introduced the time discretization $t_k = k/n,\, k = 0, \ldots, n$,  we consider the 2-dimensional Ornstein-Uhlenbeck process
\begin{equation}
  \label{eq:ou}
  X_k = \Bigl( \int_0^{t_k} e^{-\alpha_1(t_k-s)} dW^1_s ,  \int_0^{t_k} e^{-\alpha_2(t_k-s)} dW^2_s \Bigr).
\end{equation}

This Markov chain admits a useful representation as a {\it first-order auto-regressive (AR-1)}-process:
\begin{prop}\label{prop}
  For $(X_k)$ from (\ref{eq:ou}) it holds
  \begin{equation*}
    X_{k+1} = A_k X_k +T_k \epsilon_{k}, \qquad k = 0, \ldots, n\!-\! 1,
  \end{equation*}
where $A_k$ and $T_k$ are deterministic matrices and $(\epsilon_{k})$ is an i.i.d. standard normal sequence.
\end{prop}

In order to estimate the transition probabilities
\begin{equation*}
  \begin{split}
    \tp & = \Prob\bigl( X_{k} \in C_j(\gk)\, |\, X_{k-1} \in C_i(\gkm)
    \bigr)\\
& = \frac{ \Prob\bigl( X_{k} \in C_j(\gk) \cap X_{k-1} \in C_i(\gkm) \bigr)}{ \Prob\bigl( X_{k-1} \in C_i(\gkm)  \bigr)},
  \end{split}
\end{equation*}
we will therefore simulate $M$ samples of $(X_k)$ according to Proposition \ref{prop} and perform in each time-layer $k$ a Nearest Neighbor search to identify the Voronoi cell $C_j(\gk)$ in which $X_k$ falls.

Using the additional counters $p^k_{ij}$ and $p^{k}_{i}$, a serial implementation for the estimation of $\tp$ is given by Algorithm I.


\begin{algorithm}[!h]
   \begin{algorithmic}
     \FOR{$m=1, \ldots,M$} \STATE {\# { Initialization}} \STATE $x
     \leftarrow x_0,\, i \leftarrow 0,\, p_1^i \leftarrow 1$
     \FOR{$k=1, \ldots,n$} \STATE Simulate $\epsilon_k$ \STATE $x
     \leftarrow A_kx + T_k\epsilon_k$ \STATE Find NN-Index $j$ of $x$
     in $\gk$ \STATE Set \STATE $\quad p^k_{ij} \,\ +\!\!= 1$
     \STATE $\quad p^{k\!+\!1}_{j} +\!\!= 1$ \STATE $i\leftarrow j$
     \ENDFOR
     \ENDFOR
     \STATE Set $\tp \leftarrow \frac{p^k_{ij}}{p^k_{i}}, \quad
     1\leq i,j \leq N_k, 1\leq k\leq n $.\\
   \end{algorithmic}\centering
   \caption{}
   \label{fig:alg1}
 \end{algorithm}
We will adopt a numerical scenario, which has already proven  in \cite{BPW} to produce accurate results for the valuation of Swing options. Thus we set
\begin{description}
  \item[\# MC-Samples:] $M = 100.000$
  \item[\# Exercise days:] $n = 365$
  \item[Grid size:] $N = N_k = 100-500$ for $k = 1, \ldots, n$.
  \end{description}

This setting results in a computational time of $30-90$ seconds for non-parallel estimation of the transition probabilities  on a {\tt Intel Core i7} CPU@2.8GHz and $N = 100$ to $500$.

Since any parallel implementation of the above algorithm has to perform actually the following steps
\begin{enumerate}
  \item generation of the independent random numbers $\epsilon_k$
  \item a Nearest Neighbor search 
  \item updating the counters $p^k_{ij},p^k_{i}$,
\end{enumerate}
we will discuss these tasks in more detail with respect to an implementation for \cuda.

The amount of data which has to be processed in these steps when using single precision floating-point numbers is summarized in Table \ref{tab:data}.

\begin{table}[!h]
\caption{Amount of data to be processed for $N = 100-500$.}
  \centering
  \begin{tabular}{l|c|c}
    & per layer $k$ & total\\
\hline
    \# Random numbers & 100k & 36.5M\\
    \# Nearest Neighbor searches & 100k & 36.5M\\
    size of $\tp$ and $p^k_{ij}$ & 40kB - 1MB & 15 - 365MB\\
    size of grids $\gk$ & 800Byte - 4kB & 285kB - 1.5MB
  \end{tabular}
  \label{tab:data}
\end{table}

\subsection{Random number generation}\label{step1}

The challenge of random number generation on parallel devices consists in modifying the sequential random number generator algorithm in such a way, that the original sequence $\{x_n, n = 1, \ldots, M\}$ with $M = k\cdot s$
\begin{itemize}
\item is generated in independent blocks of size $s$, i.e. $k$ streams $\{x_{n\cdot s+i},\, i =  1, \ldots, s\}$, where $n = 0, \ldots, k-1$ (block approach)
\end{itemize}
or
\begin{itemize}
\item can be partitioned through a skip-ahead procedure, i.e. one generates independently $s$ streams $\{x_{n+i\cdot s},\, i = 0, \ldots, k-1\}$ for $n = 1, \ldots, s$ (skip-ahead)
\end{itemize}

The block-approach can be accomplished by generating a well chosen sequence of seed values to start the parallel computation of the random number streams.
In contrast to this,  for the skip-ahead approach we have to modify the main iteration of the random number generator itself . Nevertheless, this modification can be easily carried out for linear congruential random number generators
\[
x_{n+1} \equiv a x_n + c\!\! \mod 2^m,
\]
For this kind of generator it holds
\[
x_{n+s} \equiv A x_n + C \!\!\! \mod 2^m
\]
with $A = a^s$ and $C = \sum_{i=0}^s a^i c$.
Thus, once the coefficients $A$ and $C$ are computed, the generation of the subsequence $\{x_{n+is}, i \in \mathbb{N}\}$ is as straightforward as it is for $\{x_n, n \in \mathbb{N}\}$.

As a first parallel random number generator, we have implemented a parallel version of {\tt drand48} in \cuda, which operates in 48bit arithmetic.

A slightly more sophisticated variant of this random number generator is given by L'Ecuyer's Multiple Recursive Generator MRG32k3a (cf. \cite{MRG32k3a}) 
\begin{eqnarray*}
  x^1_n & =& (1403580\, x^1_{n-2} - 810728\, x^1_{n-3})\!\!\! \mod m_1\\
  x^2_n &=& (527612\, x^2_{n-1} - 1370589\, x^2_{n-3})\!\!\! \mod m_2\\
  x_n &=& (x^1_n - x^2_n)\!\!\! \mod m_1
\end{eqnarray*}
for $m_1 = 2^{32}-209$ and $m_2 = 2^{32}-22853$. 

Here, it is again possible to precompute constants (matrices) to generate the skip-ahead sequence
$\{x_{n+is}, i \in \mathbb{N}\}$ efficiently (see \cite{MRG32k3aP}).
An implementation in \cuda\ of this method is given by the GPU-Library of NAG.

A third kind of random number generators for \cuda\ is given by Marsaglia's XORWOW generator in the {\tt CURAND}-Library of {\tt Cuda Toolkit 3.2}.
As described in \cite{xorwow} one easily may compute starting seed values for a block approach and the random numbers sequence is then given by very small number of fast bit-shifts and XOR-operations. 
To be more precise the initialization procedure of the {\tt CURAND}-Library computes starting values for the blocks which correspond to $2^{67}$ iterations of the random engine. Moreover the main iteration of the random number generator for the state variables \texttt{v,w,x,y,z} reads

\begin{verbatim}
unsigned int curand()
{
    unsigned int t;
    t = ( x ^ (x >> 2) );
    x = y;
    y = z;
    z = w;
    w = v;
    v = (v ^ (v << 4))^(t ^ (t << 1));
    d += 362437;
    return v + d;
}
\end{verbatim}
 
To illustrate the performance of these three random number generators we have chosen a Monte Carlo simulation with a very simple integrand to illustrate the performance in simulations where the function evaluation is very cheap.
To be more precise, we estimated $\pi = 3.14159265...$ by a Monte Carlo simulation for $\frac 12 \lambda^2\bigl( B_{l^2}(0,1) \bigr)$ using $M = 10^9$ random numbers.

The results for a {\tt NVIDIA GTX 480} device and {\tt CUDA 3.2} are given in Table \ref{tab:rngs}.
The mean and the standard deviation of the MC-Estimator were computed from a sample of size 500.
\begin{table}[h!]
  \centering
  \begin{tabular}{l|c|c|c}
    RNG engine & computational time & mean & std. Dev.\\
    \hline
    drand48 & 0.2562 sec & 3.141590 & 5.2585e-05\\
    MRG32k3a & 0.2573 sec & 3.141594 & 5.20932e-05 \\
    CURAND & 0.2085 sec & 3.141592 & 5.03272e-05 \\
  \end{tabular}
\caption{Results for a Monte Carlo estimation of $\pi = 3.14159265...$}\label{tab:rngs}
\end{table}

One recognizes that the XORWOW generator from the {\tt CURAND}-library slightly outperforms the two linear congruential implementations, since  the XORWOW-step can be processed more efficiently than a modulo operation.
Nevertheless the differences between all three random number generator are rather marginal.

Especially, when we have in mind, that the original problem of swing option pricing needs only 35M random numbers in total, 
the generation of this amount of random numbers becomes negligible compared to the time spent for the nearest neighbor searches.





\subsection{Nearest Neighbor search}\label{step2}
For each MC-realization $X_k$ we have to perform a Nearest Neighbor search in every time-layer $k$ to determine the Voronoi cell $C_j(\gk)$ in which $X_k$ falls.

These Nearest Neighbor searches can be performed completely independent of each other, 
so we implemented them as sequential procedures and only have to pay attention to a proper adaption to the \cuda-compute capabilities.
%

Note here that we cannot employ the \cuda\ built-in texture fetch methods for this task, since the grids $\gk$ do in general not consist of a lattice of integer numbers.

From an asymptotical point of view, the $kd$-tree methods (cf \cite{nearestNeighbour}) obtain the fastest results for Nearest Neighbor searches of $O(\log N)$-time. 
Unfortunately, all these divide \& conquer-type approaches heavily rely on recursive function calls; a programming principle which was introduced only very recently in the \CCtwo\ specification.
Alternatively, one may implement a simple brute force Nearest Neighbor search of $O(N)$-time complexity.

The results for 36.5M NN Searches of a random number in a 2-dimensional grid can be found in Table \ref{tab:nn}.
\begin{table}[h!]
  \centering
  \begin{tabular}{c|c|c|c}
   $ N$ & brute force & $kd$-tree\\
\hline
    100  &   0.09 sec &  3.56 sec \\
250   &  0.23  sec & 5.14 sec    \\ 
500   &  0.41  sec & 6.59 sec   
    \end{tabular}
\caption{Computational time for 36.5M Nearest neighbor searches on a \FermiD\ device}
    \label{tab:nn}
\end{table}
It is striking that the brute force approach outperforms the $kd$-tree method in this setting by a huge factor, even though it suffers from a sub-optimal asymptotic behavior.

Further analysis revealed that, when using the same random number for the search in all threads of a given block,  the  $kd$-tree approach took in the same setting only 0.25 to 0.34 sec ($N = 100$ to $500$).
The dramatic slowdown of Table \ref{tab:nn}, where the NN Search is performed for different random numbers in each single thread, must be caused by a very inhomogeneous branching behavior of the single threads during the $kd$-tree traversal, which prevents the \gpgpu-scheduler of distributing the threads efficiently.

We will therefore use in the sequel the brute force approach for the further numerical experiments.



\subsection{Updating $p^k_{ij}$}
As soon as we have determined the Voronoi cells $C_i(\gkm)$ and $C_j(\gk)$ in which a realization of $(X_{k\!-\!1}, X_k)$ falls, we have to increase the counter $p^k_{ij}$.

Since, in a parallel execution of steps \ref{step1}. and \ref{step2}., it can happen that two threads try to update the same counter $p^k_{ij}$ at the same time, we arrive at the classical situation of a race condition.

Consequently, such a situation would lead to an undetermined result for the counter $p^k_{ij}$, which practically means that we randomly lose parts of the Nearest Neighbor search results.

To avoid this race condition, we are forced to employ memory locks, which are implemented in \cuda\ by means of atomic operations. Hence, we have to increment $p^k_{ij}$ by calling the \cuda-function 
\begin{verbatim}
int     atomicAdd(int* address, int val);.
\end{verbatim}

The resulting parallel procedure is stated as Algorithm II.
\begin{algorithm}[!h]
 \begin{algorithmic}
   \IF{$m=1, \ldots,M$} 
     \STATE {\# { Initialization}} 
     \STATE $x   \leftarrow x_0,\, i \leftarrow 0,\, p_1^i \leftarrow 1$ 
     \FOR{$k=1, \ldots,n$} 
        \STATE Simulate $\epsilon_k$ 
        \STATE $x \leftarrow A_kx  + T_k\epsilon_k$ 
        \STATE Find NN-Index $j$ of $x$ in $\gk$
        \STATE 
        \STATE atomic increment $p^k_{ij}$
        \STATE atomic increment $p^{k\!+\!1}_{j}$
        \STATE $i\leftarrow j$
      \ENDFOR
   \ENDIF
   \STATE Synchronize threads 
   \STATE Set in parallel $\tp \leftarrow   \frac{p^k_{ij}}{p^k_{i}}, \quad   1\leq i,j \leq N_k, 1\leq k\leq n $.\\
 \end{algorithmic}\centering
 \caption{}
 \label{fig:alg2}
\end{algorithm}

\section{Numerical results}
One of the key points in an efficient \cuda-implementation is the choice of the proper memory structure for the individual data. Table \ref{tab:memory} lists the available memory types in \CCone.

\begin{table}[!h]
  \centering

\begin{tabular}{|*{3}{l|}}
\hline
local memory & {\bf not} cached & 16kB per thread  \\

constant memory  & cached  & 64kB per device  \\
shared memory &  n/a &  16kB per block \\
global memory &  {\bf not} cached & $\approx 1$GB per device\\ \hline
\end{tabular}
  \caption{Memory types for \cuda\ compute capability 1.x}
  \label{tab:memory}
\end{table}
Note that {\it shared memory} is (beneath the processor registers) the fastest memory available in \cuda, since it resides very close to the processor cores. 
There are 16kB of shared memory available per Multiprocessor, whose content is read- and writable by any thread in the same block of a grid.

The other memory types in Table \ref{tab:memory} are about 400 times slower than shared memory except {\it constant memory} which is cached and therefore achieves a similar read performance as shared memory.

Taking into account the sizes of the arrays $\tp, p^k_{ij}$ and $\gk$ from Table \ref{tab:data}, there is no other possibility for the above algorithm than to place all the arrays in global memory, since any thread has to access the arrays  $\tp, p^k_{ij}$ and $\gk$ for any $k,\, 1\leq k \leq n$.

The fact that these arrays have to reside in global memory especially slows down the Nearest Neighbor searches, which rely on a fast access to the grid points of $\gk$.

We therefore present another approach, which maximizes the parallel execution by splitting up the problem into smaller parts, that can make use of faster memory.

Note that due to Proposition \ref{prop} we can directly simulate the couple $(X_k, \varepsilon_k)$ in order to get a realization of $(X_k, X_{k+\!1})$ without the need of generating $X_l, \, l < k$.

Thus, if we accept to generate twice the amount of random numbers and double the number of Nearest Neighbor searches, we arrive at Algorithm III.

\begin{algorithm}[!h]
  
\begin{algorithmic}
  \IF{$k=1, \ldots,n$} 
     \IF{$m=1, \ldots,M$} 
        \STATE Simulate $X_k,  \epsilon_k$ 
        \STATE 
        \STATE Find NN-Index $i$ of $X_k$ in $\gk$
        \STATE Find NN-Index $j$ of $A_k X_k + T_k \epsilon_k$ in  $\gkk$ 
        \STATE 
        \STATE atomic increment  $p^k_{ij}$
        \STATE atomic increment $p^{k}_{i}$
      \ENDIF
      \STATE Synchronize 
      \STATE Set in parallel $\tp \leftarrow  \frac{p^k_{ij}}{p^k_{i}}, \quad 1\leq i,j \leq N_k$
  \ENDIF
\end{algorithmic}\centering
\caption{}
\label{fig:alg3}
\end{algorithm}
Here, we do not only parallelize with respect to the MC-samples (pathwise), but also with respect to the time-layer $k$.
Therefore, we are able to perform the whole MC-simulation of a given time-layer $k$ (i.e. the inner loop) on a single Multiprocessor (i.e. within a single block in \cuda-terminology).

Hence, we can store the involved grids $\gk$ and $\gkk$ entirely in shared memory and benefit from a huge performance gain.

\begin{figure}[!t]
   \centering
   \includegraphics[width=0.8\textwidth]{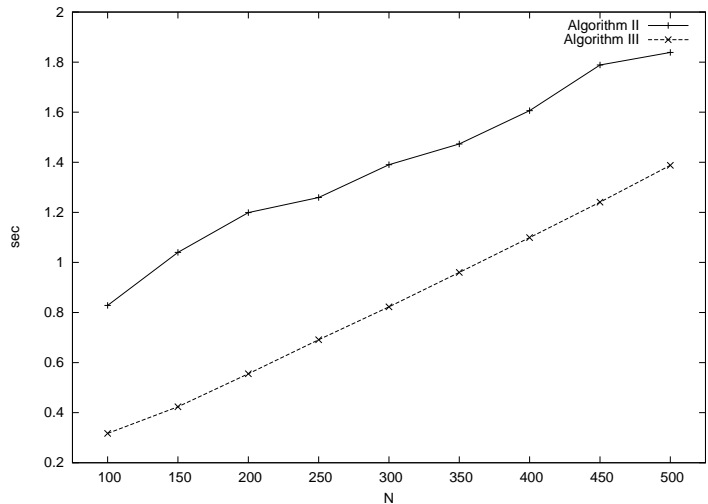}
   \caption{Linear performance of Algorithms II \& III with respect to $N$ on a \oldD\ device}
   \label{fig:plot}
\end{figure}

This can be seen in Table \ref{tab:results} and Figure \ref{fig:plot}, which demonstrates that the shared memory implementation - performing even twice as many Nearest Neighbor searches - is still significantly faster than the usual pathwise parallelization for \CCone.

\begin{table}[!h]
  \centering

  \begin{tabular}{*{4}{c|}}
$N$ & $100$ & $250$  & $500$ \\
\hline
Algorithm II &  $0.82$ sec &  $1.25$ sec & $1.83$ sec\\
Algorithm III &  $0.31$ sec &  $0.68$ sec & $1.38$ sec
\end{tabular}
  \caption{Computational times for the transition probabilities on a {\tt NVIDIA GTX 295} device}
  \label{tab:results}
\end{table}

All the computations for \CCone\ were performed on a {\tt NVIDIA GTX 295} GPGPU, \cuda\ Toolkit 2.3 and \nvidia\ X-Driver 190.53 for 64bit Linux. 
The running times in Table \ref{tab:results} also include the transfer of the transition probabilities $\tp$ back to the host CPU.

Furthermore, we have chosen in all examples 256 - 512 threads per block and overall 365 - 400 blocks. This choice was optimal for our setting.
Note here, that the shared memory algorithm performs $73 \cdot10^6$ Nearest Neighbor searches in a 2-dimensional grid.
Assuming that the brute force Nearest Neighbor search
\[
\bigl( \min < (x_1-y_1)^2 + (x_2-y_2)^2\bigr)
\]
for each grid point is equivalent to 6 FP-operations (3 additions, 2 multiplications, 1 comparison), we already arrive for $N=500$ at a computing power of approx. 175 GFLOPS only for the Nearest Neighbor searches (the pure kernel execution takes in this case 1.25sec).
Compared to the peak performance of 895 GFLOPS for one unit in the {\tt NVIDIA GTX 295}-device, this fact underlines that our implementation exploits a great amount of the theoretically available computing power of a \gpgpu-devices.

\subsection{Progress in hardware: the Fermi-architecture}
\begin{figure}[!t]
   \centering
   \includegraphics[width=0.8\textwidth]{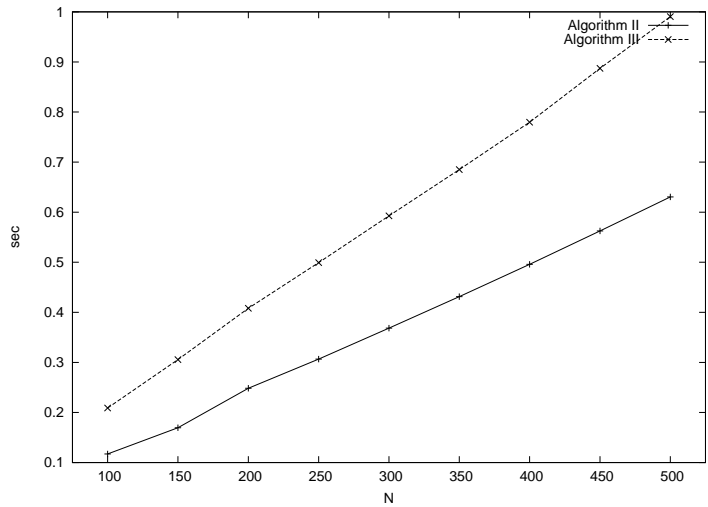}
   \caption{Performance of Algorithms II \& III with respect to $N$ on a \FermiD\ device}
   \label{fig:plotFermi}
\end{figure}

With the arrival of \CCtwo\ and the Fermi-architecture, there are now L1- and L2 caches available of up to 48kB per block. It turned out that this change in hardware design has strong implications on the performance of Algorithm \ref{fig:alg2}. 
As it can be seen in Table \ref{tab:resultsFermi} and Figure \ref{fig:plotFermi}, 
\begin{table}[!h]
  \centering
  \begin{tabular}{*{4}{c|}}
$N$ & $100$ & $250$  & $500$ \\
\hline
Algorithm II &  $0.11$ sec &  $0.30$ sec & $0.63$ sec \\
Algorithm III &  $0.21$ sec &  $0.50$ sec & $0.99$ sec\\
\end{tabular}
  \caption{Computational times for the transition probabilities on a \FermiD\ device}
  \label{tab:resultsFermi}
\end{table}
the new cache can nearly completely compensate the advantage of the shared memory usage in Algorithm \ref{fig:alg3}.
Moreover, both parallelizations differ roughly by a factor of two which is caused by the fact that algorithm \ref{fig:alg3} has to perform twice the number of Nearest Neighbor searches than Algorithm \ref{fig:alg2}.
The computations for \CCtwo\ were performed on a \FermiD\ GPGPU, \cuda\ Toolkit 3.2 and \nvidia\ X-Driver 260.19.29 for 64bit Linux

\section{Conclusion}
We have shown in this paper that the use of \gpgpu-devices is quite efficient for the estimation of transition probabilities in a Quantization Tree. 
Although we resorted for the Nearest Neighbor search, which is the most compute intensive part of the algorithm, to the sub-optimal brute-force approach, we could achieve by means of the massive computing power of a \gpgpu-device a speed-up of factor 200 compared to a serial CPU implementation.
Those implementations can therefore be used for online estimation of the transition probabilities in time-critical applications in practice, which is not possible for a CPU implementation that can take more than 1 min for the same task.


\section*{Acknowledgment}
The authors would like to thank J. Portes for setting up machines and NAG for providing the \cuda\ routines for the MRG32k3a generator.


%% file: tree_t
\begin{picture}(0,0)%
\includegraphics{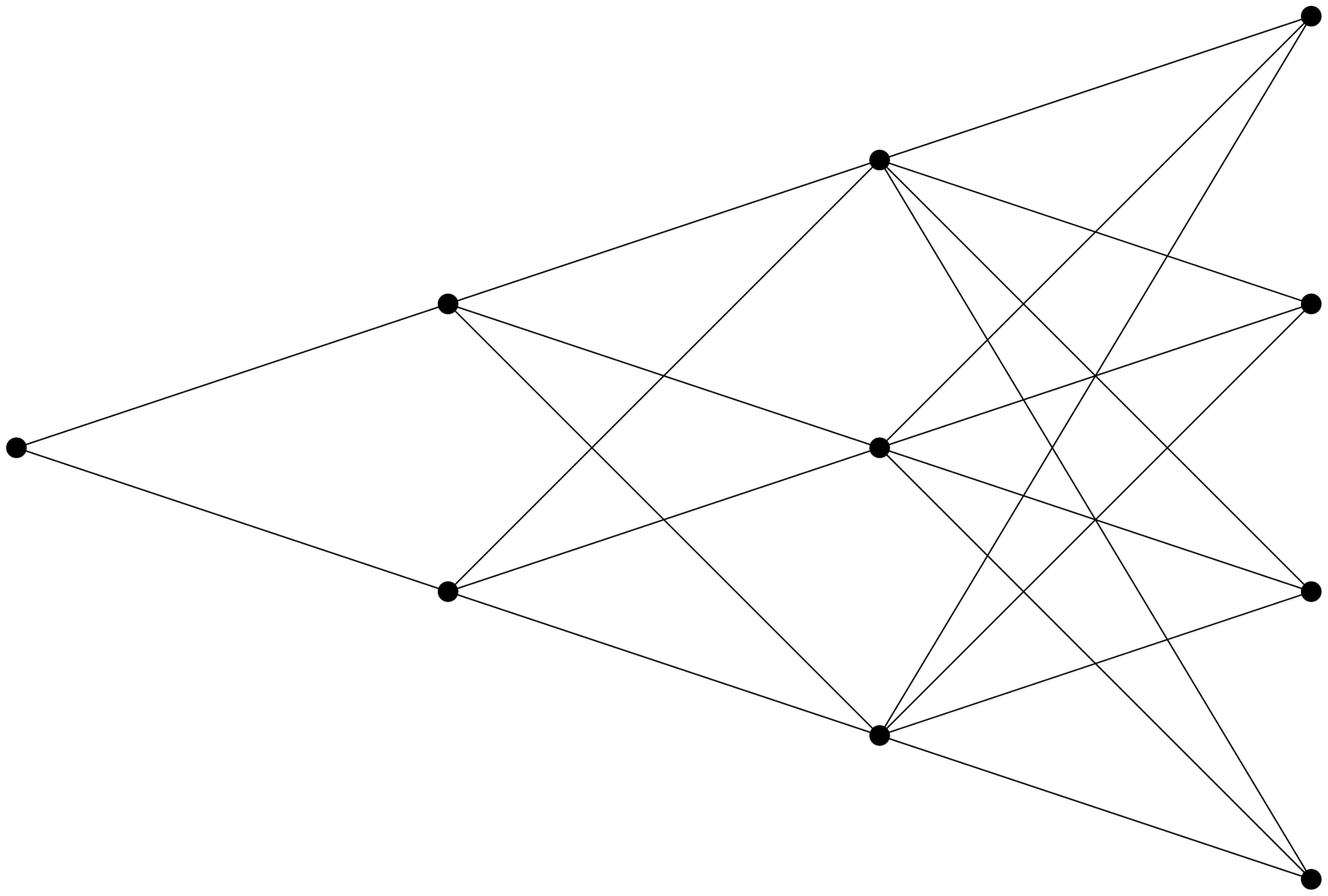}%
\end{picture}%
\setlength{\unitlength}{4144sp}%
\begingroup\makeatletter\ifx\SetFigFont\undefined%
\gdef\SetFigFont#1#2#3#4#5{%
  \reset@font\fontsize{#1}{#2pt}%
  \fontfamily{#3}\fontseries{#4}\fontshape{#5}%
  \selectfont}%
\fi\endgroup%
\begin{picture}(8850,6373)(346,-7109)
\put(1891,-2941){\makebox(0,0)[lb]{\smash{{\SetFigFont{25}{30.0}{\familydefault}{\mddefault}{\updefault}{\color[rgb]{0,0,0}$\pi^1_{1,1}$}%
}}}}
\put(1891,-4561){\makebox(0,0)[lb]{\smash{{\SetFigFont{25}{30.0}{\familydefault}{\mddefault}{\updefault}{\color[rgb]{0,0,0}$\pi^1_{1,2}$}%
}}}}
\put(9181,-2851){\makebox(0,0)[lb]{\smash{{\SetFigFont{25}{30.0}{\familydefault}{\mddefault}{\updefault}{\color[rgb]{0,0,0}$x^3_2$}%
}}}}
\put(9181,-4651){\makebox(0,0)[lb]{\smash{{\SetFigFont{25}{30.0}{\familydefault}{\mddefault}{\updefault}{\color[rgb]{0,0,0}$x^3_3$}%
}}}}
\put(9181,-6451){\makebox(0,0)[lb]{\smash{{\SetFigFont{25}{30.0}{\familydefault}{\mddefault}{\updefault}{\color[rgb]{0,0,0}$x^3_4$}%
}}}}
\put(361,-3751){\makebox(0,0)[lb]{\smash{{\SetFigFont{25}{30.0}{\familydefault}{\mddefault}{\updefault}{\color[rgb]{0,0,0}$x^0_1$}%
}}}}
\put(9181,-1051){\makebox(0,0)[lb]{\smash{{\SetFigFont{25}{30.0}{\familydefault}{\mddefault}{\updefault}{\color[rgb]{0,0,0}$x^3_1$}%
}}}}
\put(6121,-6991){\makebox(0,0)[lb]{\smash{{\SetFigFont{25}{30.0}{\familydefault}{\mddefault}{\updefault}{\color[rgb]{0,0,0}$2$}%
}}}}
\put(361,-6991){\makebox(0,0)[lb]{\smash{{\SetFigFont{25}{30.0}{\familydefault}{\mddefault}{\updefault}{\color[rgb]{0,0,0}$k=0$}%
}}}}
\put(8911,-6991){\makebox(0,0)[lb]{\smash{{\SetFigFont{25}{30.0}{\familydefault}{\mddefault}{\updefault}{\color[rgb]{0,0,0}$3$}%
}}}}
\put(3511,-6991){\makebox(0,0)[lb]{\smash{{\SetFigFont{25}{30.0}{\familydefault}{\mddefault}{\updefault}{\color[rgb]{0,0,0}$1$}%
}}}}
\end{picture}%